\documentclass[journal]{IEEEtran}

\ifCLASSINFOpdf
\else
   \usepackage[dvips]{graphicx}
\fi
\usepackage{url}

\usepackage{pbox}
\usepackage{tikz}
\usepackage{pgfgantt}
\usepackage{subfigure}
\usepackage{gensymb}
\usepackage{amsmath}
\usepackage{adjustbox} 
\usepackage{amsfonts}
\usepackage{graphicx}
\usepackage{array}
\usepackage{multirow}
\usepackage{makecell}
\usepackage{enumitem}
\usepackage{upgreek}
\usepackage{graphicx}
\usepackage{epstopdf}

\usepackage{soul}

\usepackage{algorithm}
\usepackage[noend]{algpseudocode}

\usetikzlibrary{matrix,shapes,arrows,positioning,chains}
\tikzstyle{process} = [rectangle, minimum width=3cm, minimum height=1cm, text centered, draw=black]
\tikzstyle{arrow} = [thick,->,>=stealth]
\tikzstyle{io} = [trapezium, trapezium left angle=70, trapezium right angle=110, text centered]

\begin{document}

\title{Evolution of urban areas and land surface temperature}

\author{Sudipan Saha,  Tushar Verma, Dario Augusto Borges Oliveira
\thanks{Sudipan Saha is with Yardi School of Artificial Intelligence, Indian Institute of Technology Delhi, New Delhi, India. E-mail: sudipan.saha@scai.iitd.ac.in}
\thanks{Tushar Verma is with Yardi School of Artificial Intelligence, Indian Institute of Technology Delhi, New Delhi, India. E-mail: aiy227514@scai.iitd.ac.in}
\thanks{Dario Augusto Borges Oliveira is with the School of Applied Mathematics, Getulio Vargas Foundation, Brazil. E-mail: darioaugusto@gmail.com}}

\markboth{}
{Shell \MakeLowercase{\textit{et al.}}: Bare Demo of IEEEtran.cls for IEEE Journals}
\maketitle

\begin{abstract}

With the global population on the rise, our cities have been expanding to accommodate the growing number of people. The expansion of cities generally leads to the engulfment of peripheral areas. However, such expansion of urban areas is likely to cause increment in areas with increased land surface temperature (LST).
By considering each summer as a data point, we form LST multi-year time-series and cluster it to obtain spatio-temporal pattern. We observe several interesting phenomena from these patterns, e.g., some clusters show reasonable similarity to the built-up area, whereas the locations with high temporal variation are seen more in the peripheral areas. Furthermore, the LST center of mass shifts over the years for cities with development activities tilted towards a direction. We conduct the above-mentioned studies for three different cities in three different continents.  
\end{abstract}


\IEEEpeerreviewmaketitle

\fboxsep=0mm
\fboxrule=0.1pt

\section{Introduction}

The world has been urbanizing rapidly in the last decades. However, such expansion of urban areas contributes significantly to the climate change  \cite{carvalho2017urban}. One most prominent effect of climate change is the increase in land surface temperature (LST).  Surfaces of impermeable materials reduce evapotranspiration \cite{halder2021evaluation,stewart2012local}, thus contributing to the increase in LST.
Increases in urban temperature and intensity of Urban Heat Island (UHI) are easily amplified by human activities \cite{house2011advances}. Along with increasing urbanization, heat waves have become  more frequent and intense in the last decades \cite{rohini2016variability}, causing increased mortality.
\par 
Owing to the availability of several freely available Earth observation sensors, it is now possible to monitor the urban changes at high spatial and temporal resolution. As an example, Sentinel-2 images can be used to generate building maps at up to 10 meter/pixel spatial resolution. On the other hand, several available sensors provide information related to LST at reasonable spatial resolution. As an example, LST can be estimated from Moderate Resolution Imaging Spectroradiometer (MODIS) \cite{yoo2018estimation} or Landsat-8 thermal bands  \cite{rajeshwari2014estimation}. This, in turn, makes it possible to study  the relationship between land surface changes and LST pattern at a high spatial resolution.
\par
Remote sensing time series data offers several advantages and can provide more comprehensive and detailed insights compared to analyzing single-time data. The temporal dimension provides an opportunity to analyze trends and variations over time, thus helping to understand the dynamics of various phenomena. Furthermore, time-series data enables us to identify long-term trends that are not apparent in single-time snapshots. Following this cue, in this work we investigated 20 years LST summer time-series data.  Recent works have shown the potential to discover interesting spatial patterns directly from unlabeled Earth observation data \cite{saha2022unsupervised, pan2022hierarchical}. In a similar manner, we cluster the LST time-series to obtain different clusters/groups from the analyzed urban scene. We furthermore analyze whether such clusters confirm the hypothesis that there is a strong relationship between urban expansion and increasing LST. Instead of focusing on absolute value of LST, our work focuses on relative changes of LST over the years. Furthermore, to understand the differences in such LST dynamics among different parts of the world, we conducted studies on three cities, each in different countries/continents: Kolkata (India), Sao Paulo (Brazil), and Munich (Germany). Two of the three studied cities are close to tropics.
\par
The key contributions of this paper are as follows:
\begin{enumerate}
\item We analyze LST summer time-series of several years for three particular cities. The number of years are chosen in a manner such that it helps us to comprehend the dynamics occurring within the boundaries of our memorable history.
\item We apply unsupervised clustering to find spatio-temporal patterns from LST time-series. We also find the locations showing high temporal variation.
\item We furthermore study the relationship of the above-mentioned patterns with built-up area pattern.
We study the similarity in pattern between the growth in urban built-up areas and the areas with an increase in LST.
\item We also compute LST center of mass and its shift over time.
\item We demonstrate the above studies in the context of three cities, Kolkata, Sao Paulo, and Munich.
\end{enumerate}
\par

We organize the rest of the paper as follows. A few prior works that studied urbanization and LST are discussed in Section \ref{sectionRelatedWorks}. Section \ref{sectionProposedMethod} outlines the data sources and methodologies used in the study.
Specific analyzed scenes and experimental results are detailed in Section \ref{sectionExperimentalResult}. Finally, we conclude the paper in Section \ref{sectionConclusion}.

\section{Urbanization and LST}
\label{sectionRelatedWorks}
Several previous works have studied the different aspects related to urbanization and land surface temperature. The difference in  temperature between urban and adjacent less-developed rural areas has been noted in studies related to
Urban Heat Island (UHI) \cite{mirzaei2015recent}. Several works have applied satellite images for UHI related studies \cite{kaur2022spatial,garcia2022analysis}. Most studies employ MODIS data for LST estimation \cite{yoo2018estimation,phan2018application}. In \cite{yoo2018estimation}, daily maximum and minimum air temperatures for two cities in South Korea are estimated using MODIS LST time series. Urban heat island from 2015-2019  is  extracted using Landsat 8 data studied in \cite{kaur2022spatial} for a district in Punjab, India. Their study showed that the higher-temperature pixels lie in urban areas with dense infrastructure. More recently, Sentinel-3 data has also been used for UHI-related studies, e.g., \cite{garcia2022analysis} showed that LST and surface UHI are statistically related and such relation intensifies under heat wave conditions. In \cite{halder2022investigating}, a positive correlation between increment in the built-up land and the average temperature increase is shown for Seville, Spain. Such a relationship between urban expansion and urban heat island expansion is also shown in \cite{chunmei2022urban,huang2022detecting}. Some works have attempted to downscale LST data to improve spatial resolution \cite{sismanidis2016assessing}. Similar to \cite{halder2022investigating}, we also show a relationship between the increment in the built-up area and the increase in temperature. However, we demonstrate this phenomenon in the context of the peripheral areas that have been recently urbanized. Furthermore, we also perform  clustering on LST time-series to get a better understanding of the semantics of different urban areas.

\section{Data Sources and Methods}
\label{sectionProposedMethod}
Taking a series of LST maps computed over same geographic location, however at different time instances, we form a time series and apply clustering on the LST time-series to obtain segmentation maps. Furthermore, we estimate built-up area maps and built-up area change map.
\par
\textbf{LST:} A few satellite-based sensors, e.g., MODIS  allow us to estimate LST from specific bands \cite{wan2006modis,wan2008new}.
The MODIS's LST product is a valuable information source about the Earth surface's thermal characteristics at a global scale. It captures radiation in two thermal bands and exploits their spectral differences to correct for the atmospheric effect. The final product is LST in Kelvin that can be easily converted to degree Celsius. LST difference map between two time instances can be generated as the difference in LST between those time instances.
\par
\textbf{LST time-series clustering:} 
We form LST time-series $X$ consisting of $T$ images over a scene of spatial dimension of $R\times C$. Each image corresponds to one year.
In other words, each pixel in the analyzed scene $X$ is represented by a vector of length $T$. Since we have used 20 years data, $T=20$ in this case. Following this, we cluster the pixels into different groups/clusters using k-Means clustering method. The objective of the clustering is to find the group of pixels that share similar pattern w.r.t. LST over the analyzed years. In this study, we set $k=4$, as we empirically found that in the analyzed urban areas built-up areas decompose into two LST groups. Similarly, we assume two groups for areas that are not built-up and thus set $k=4$.
While any other similar value (e.g., $k=3$) could be set, this would not change the gist of this work.
\par
\textbf{Locations with high temporal variation:} Spatial locations with high temporal variation are of interest due to the uncertainty associated with such locations. We compute pixelwise standard deviation along the temporal dimension. Following this, we apply Otsu's threshold \cite{otsu1979threshold} to segregate regions with higher variation from those with lower variation.
The region with higher variance need not have any relationship with the urban core area, as they may be consistently warm and show less variation. Rather, the areas that are seeing infrastructure development more recently are more likely to  fall in this group.
\par
\textbf{LST center of mass:} 
The center of mass is generally defined as an imaginary point in a body of matter where the total weight of the body is thought to be concentrated. This concept has been used in some works to characterize the shift of urban expansion \cite{yu2021spatiotemporal}. Following the same concept, we compute LST center of mass for each year. We first estimate a threshold on the latest available year using Otsu's threshold \cite{otsu1979threshold} and use this threshold to binarize each LST images. Following this we compute the center of mass on these binary images. We hypothesize that the shift of LST center of mass is also an indicator of nature of urban expansion. We envision three scenarios:
\begin{itemize}
    \item If urban expansion is tilted towards a side of the city, this will in turn cause LST to increase there, while it will remain fairly constant in the other places. This will cause the LST center of mass  to drift towards that side.
    \item If the urban expansion happens uniformly in all directions around the city, then there would not be much shift of the LST center of mass.
    \item If the urban expansion is negligible, then there would not be much shift of the LST center of mass.
\end{itemize}
\par
\textbf{Similarity computation between LST-based clusters and built-up density:} Over the analyzed scenes, we compared the LST clusters to the building density map derived from Sentinel-2 images. This is to verify our hypothesis that the temperature-wise top clusters obtained from LST time-series clustering correspond to the urban built-up area. 
\par
Using NIR (band 8) and SWIR (band 11), we first compute  Normalized Difference Built-up Index (NDBI) that is commonly used for estimating impervious urban areas \cite{kuc2019sentinel}. We further binarize grayscale NDBI images using Otsu's threshold \cite{otsu1979threshold} to classify each pixel as either built-up or not. Following this, to remove noisy built-up pixels, we apply smoothing (mean) filter and further binarize it. The final outcome is a binarized built-up density map indicating presence of built-up area in the surrounding. 
\par
To compare the similarity between LST-based cluster and built-up density, we simply compute Intersection-over-Union between between the target clusters and a binarized built-up density map.  
W.r.t. above-mentioned assumption about relationship between temperature-wise top clusters and built-up area, we note that while LST captures the temporal dynamics as well, the built-up density map is static and does not capture temporal dynamics. This may cause some dissimilarity between them.

\begin{figure*}[!h]
\centering
{Min\hskip0.5em\includegraphics[height=0.5 cm, width = 3 cm]{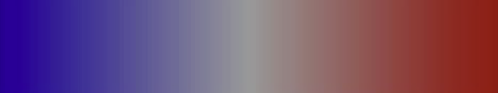}\hskip0.5em Max \quad Color legend for LST variation} 
\\
\subfigure[]{%
            
         \fbox{\includegraphics[height=4.2 cm]{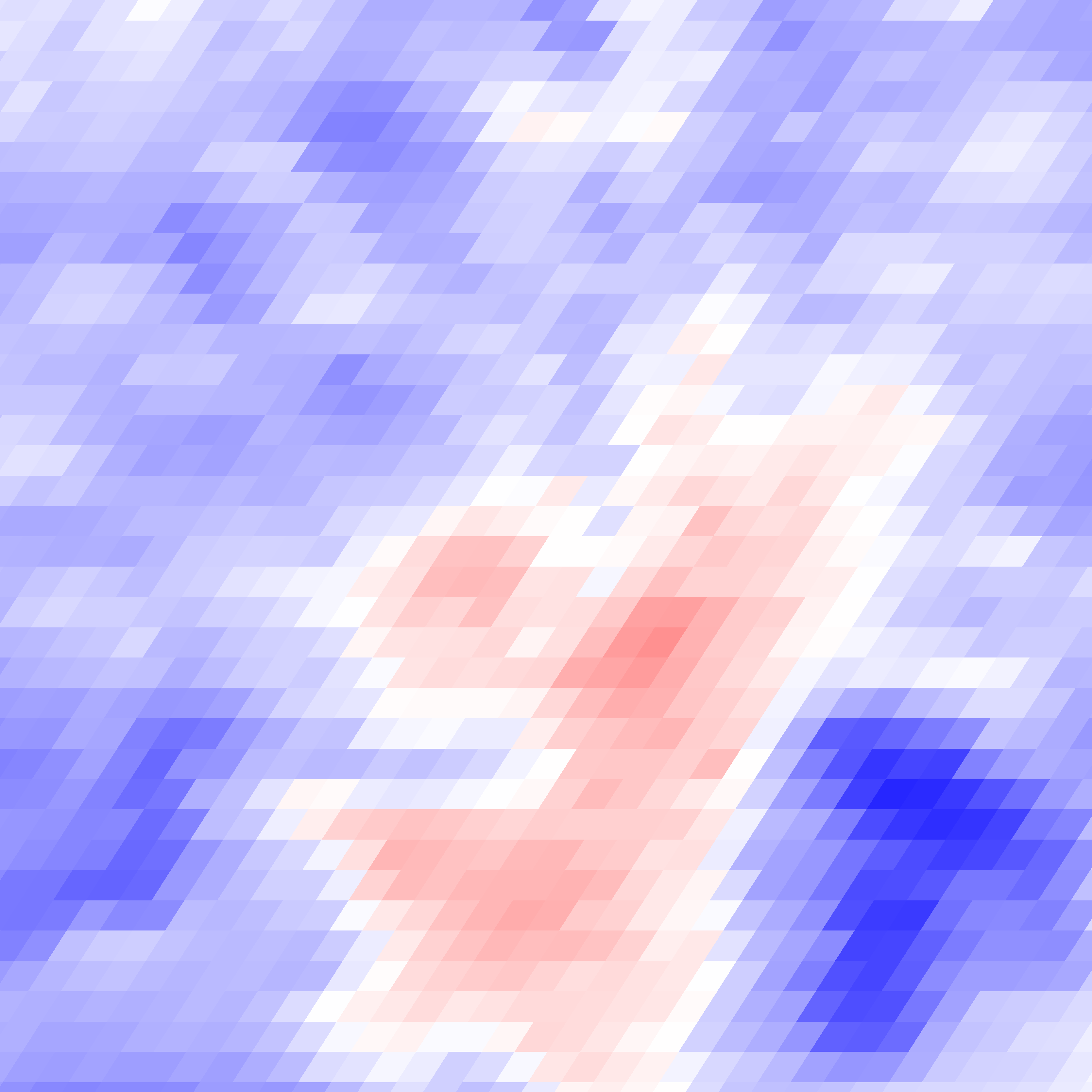}}
            \label{figureKolkata2004}
        }%
\hspace{0.2 cm}
\subfigure[]{%
            
         \fbox{\includegraphics[height=4.2 cm]{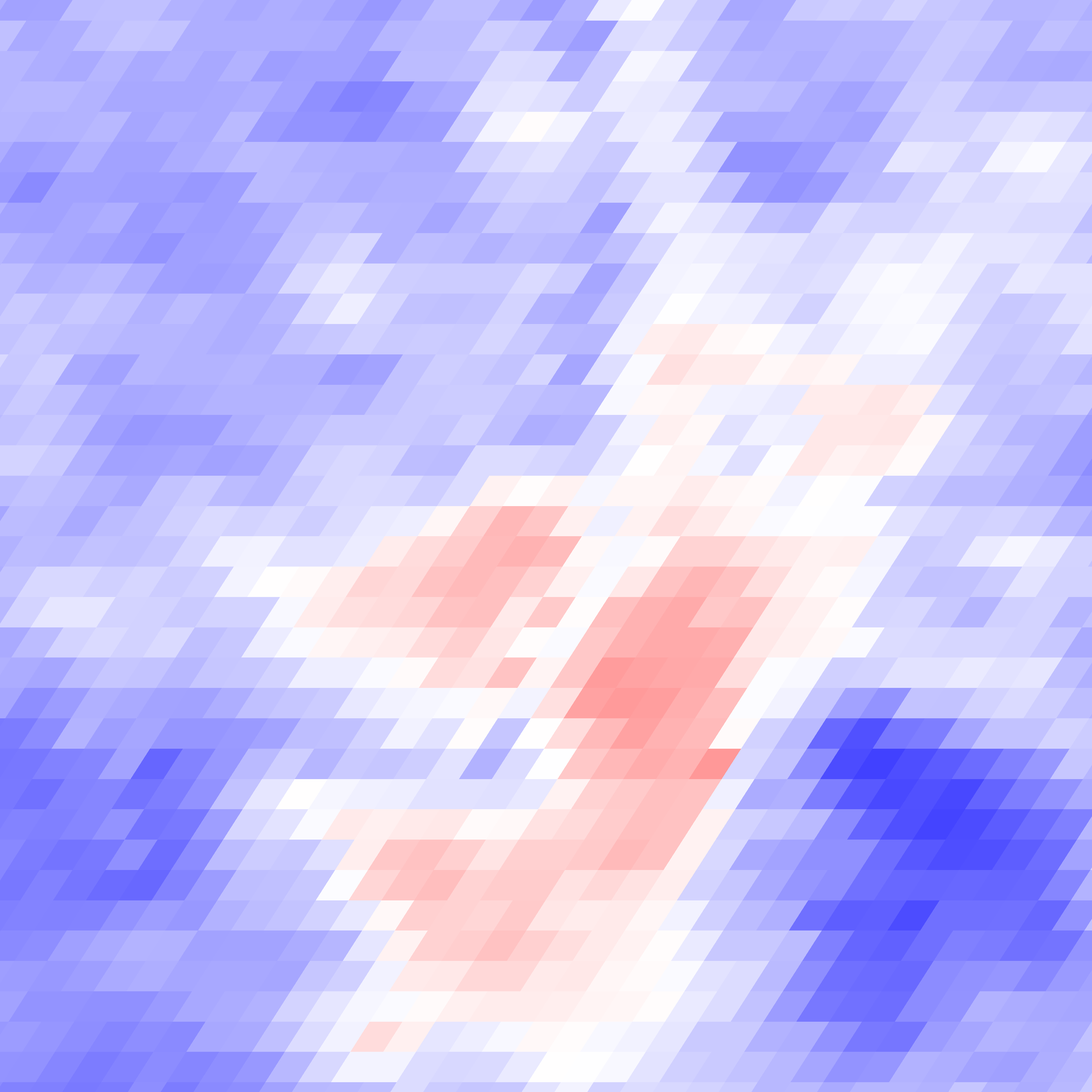}}
            \label{figureKolkata2005}
        }%
\hspace{0.2 cm}
  \subfigure[]{%
 \fbox{\includegraphics[height=4.2 cm]{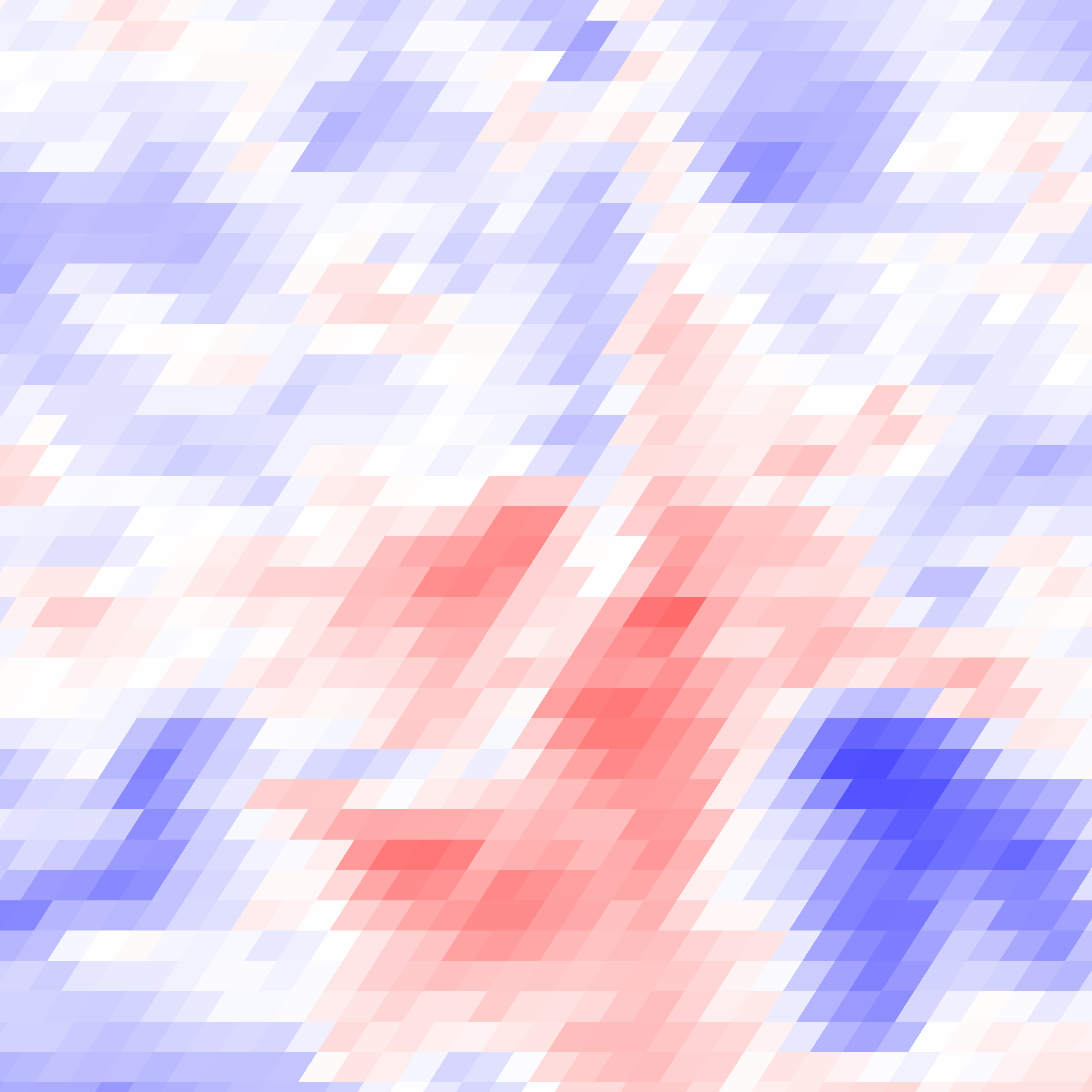}}
            \label{figureKolkata2006}
        }%
\\

\subfigure[]{%
            
         \fbox{\includegraphics[height=4.2 cm]{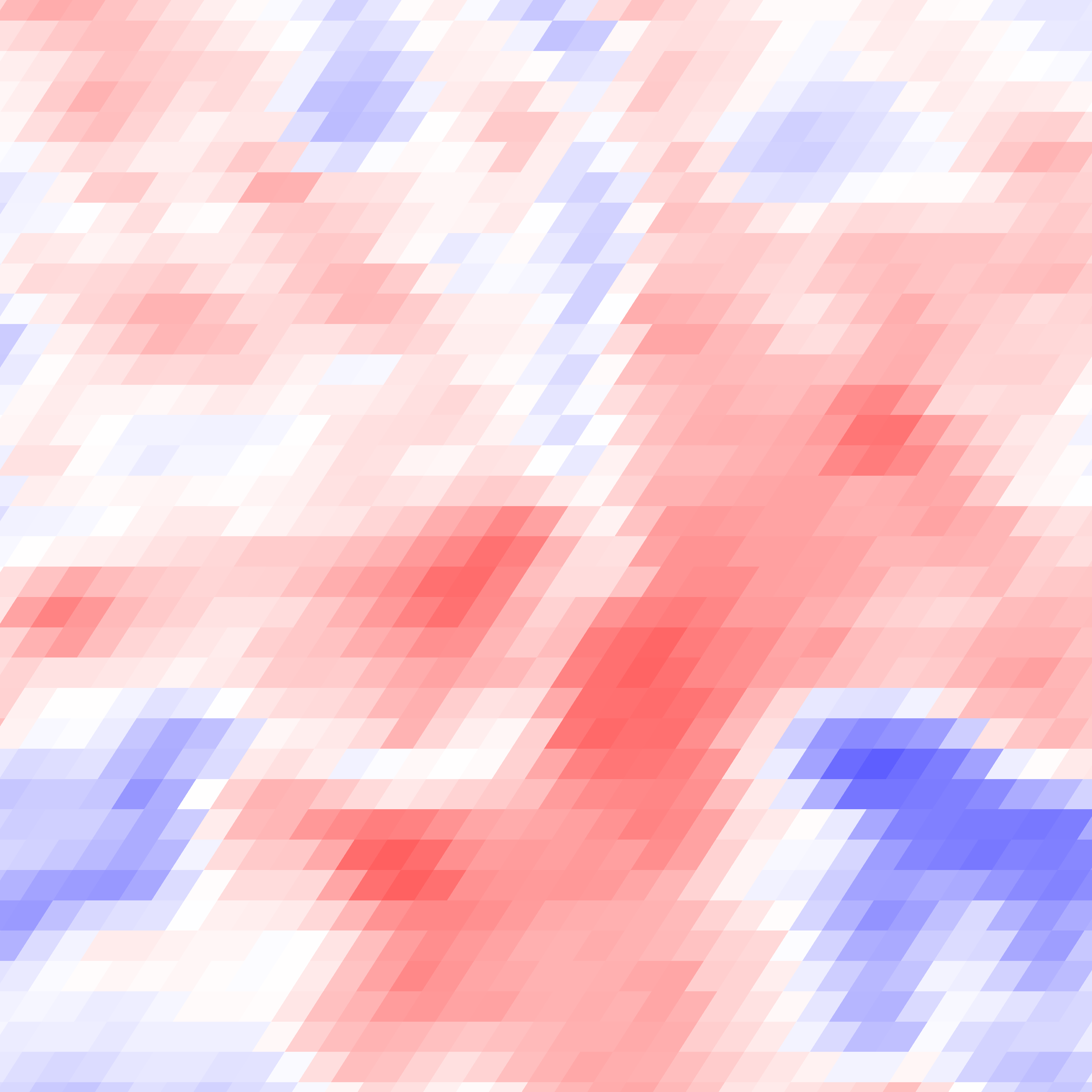}}
            \label{figureKolkata2021}
        }%
\hspace{0.2 cm}
\subfigure[]{%
            
         \fbox{\includegraphics[height=4.2 cm]{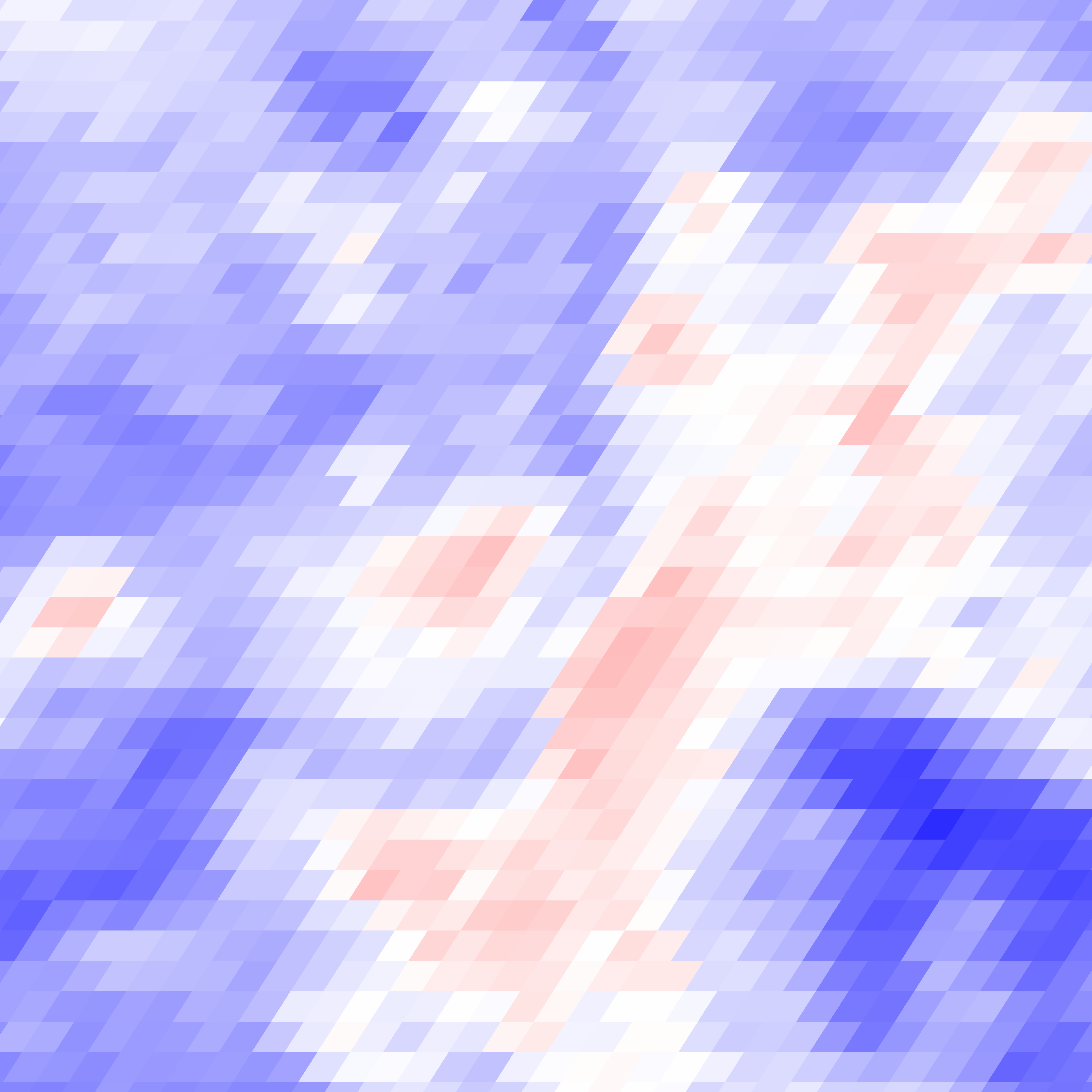}}
            \label{figureKolkata2022}
        }%
\hspace{0.2 cm}
  \subfigure[]{%
 \fbox{\includegraphics[height=4.2 cm]{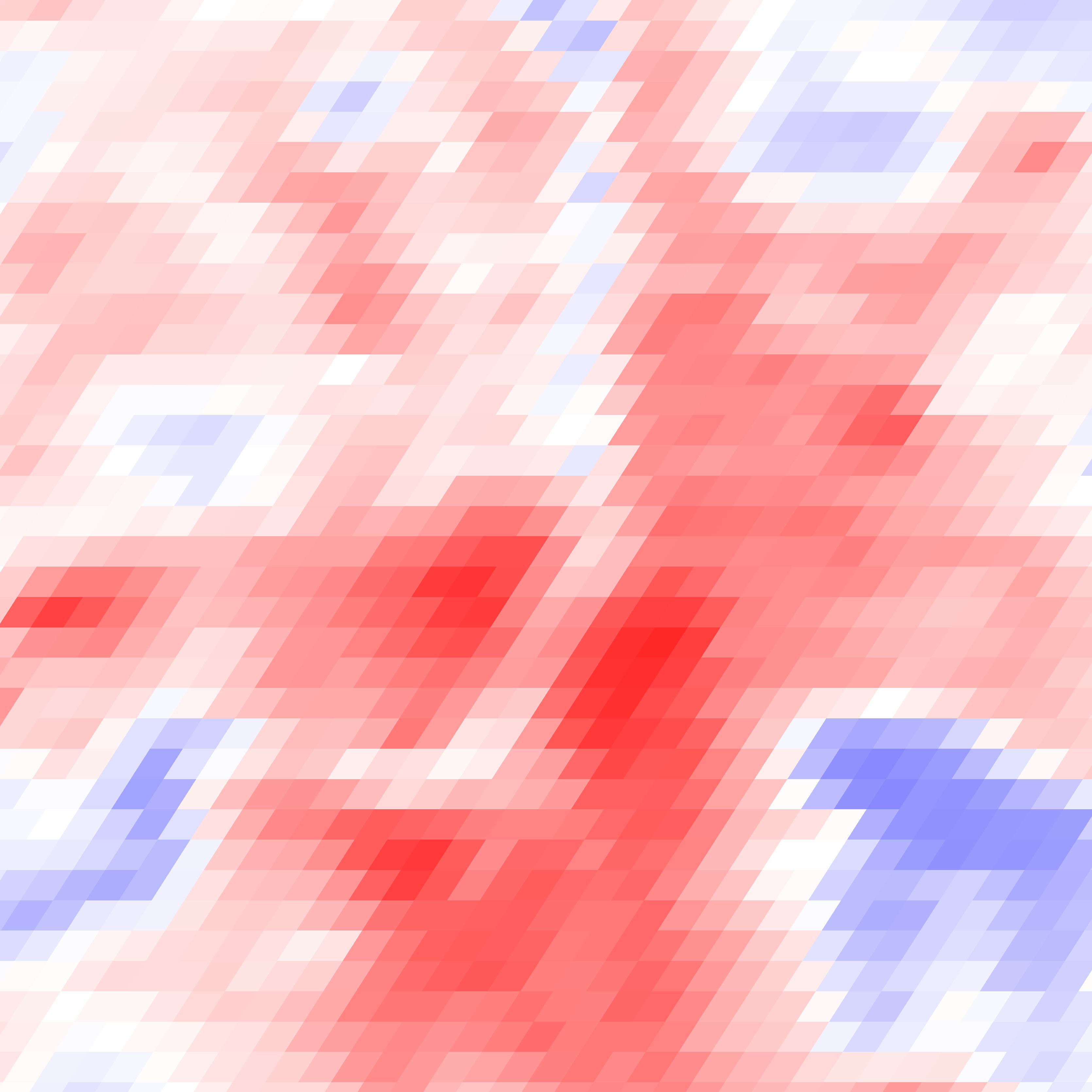}}
            \label{figureKolkata2023}
        }%
\\
\subfigure[]{%
            
         \fbox{\includegraphics[height=4.2 cm]{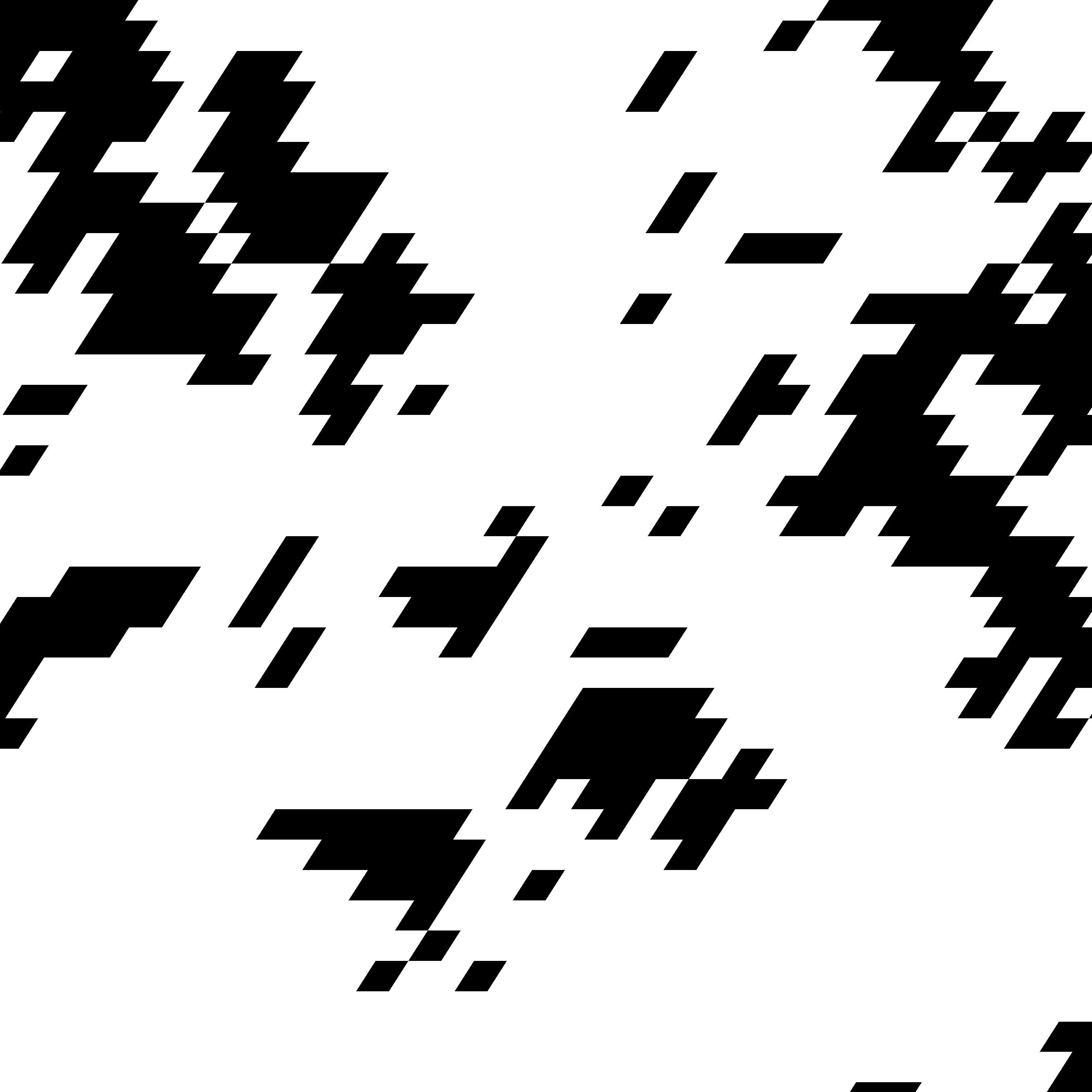}}
            \label{figureKolkataStd}
        }%
\hspace{0.2 cm}
\subfigure[]{%
            
         \fbox{\includegraphics[height=4.2 cm]{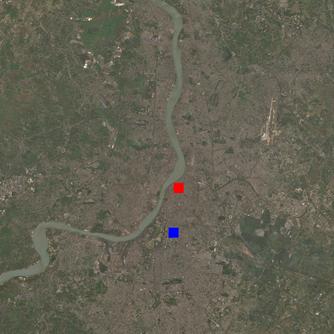}}
            \label{figureKolkataRGB}
        }%
\hspace{0.2 cm}
\subfigure[]{%
            
         \fbox{\includegraphics[height=4.2 cm]{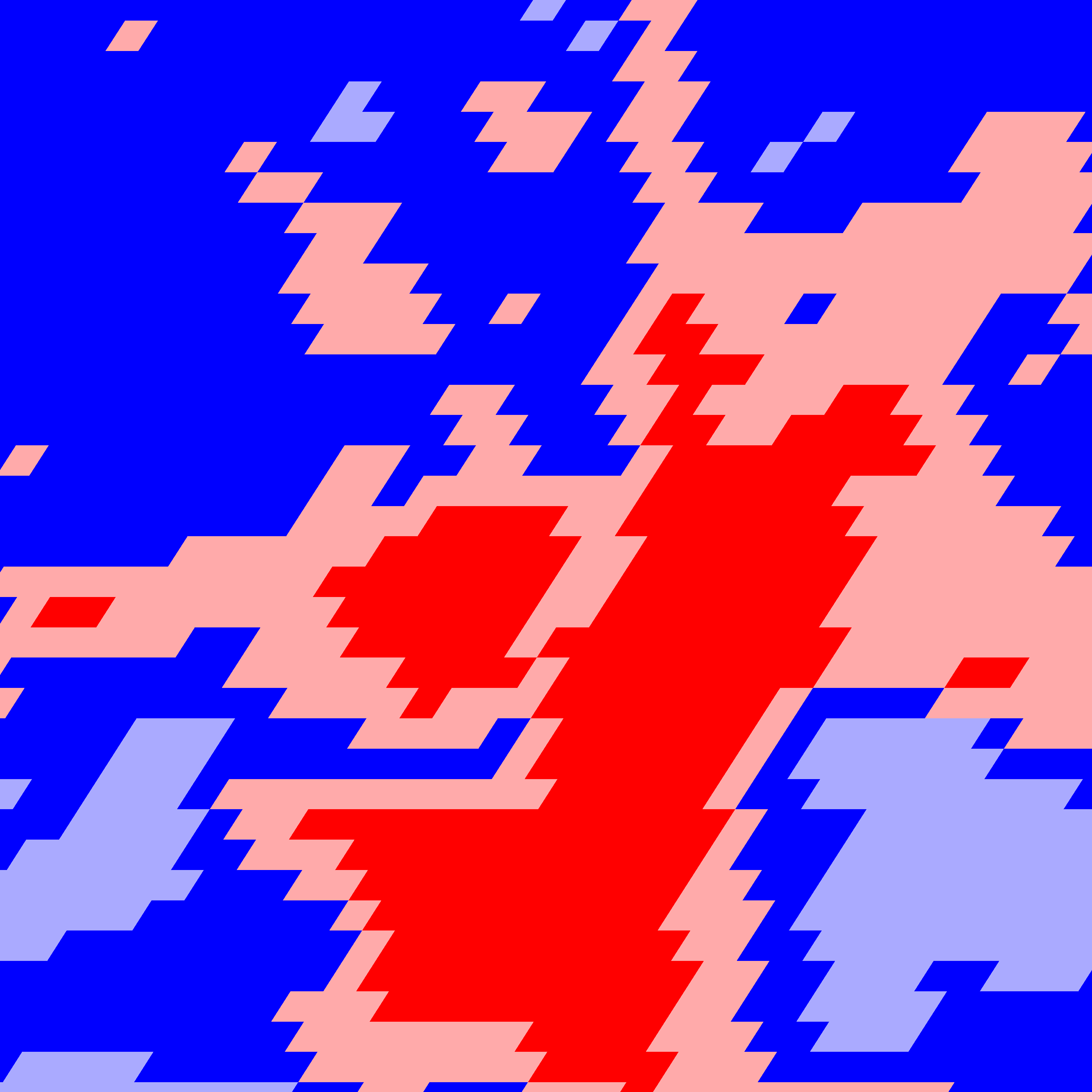}}
            \label{figureKolkataCluster}
        }%

\caption{Visualization of LST variation in a ROI around Kolkata, India for year: (a) 2004, (b) 2005, (c)  2006, (d) 2021, (e) 2022, (f) 2023. (g) Areas with strong LST temporal variation shown in black. (h) RGB image of the analyzed scene with center of mass in 2004 and 2023 shown in blue and red, respectively. (h) clusters/groups obtained from LST time-series.}
\label{figureLSTKolkata}
\end{figure*}

\begin{figure}[!h]
\centering
\subfigure[]{%
            
         \fbox{\includegraphics[height=2.2 cm]{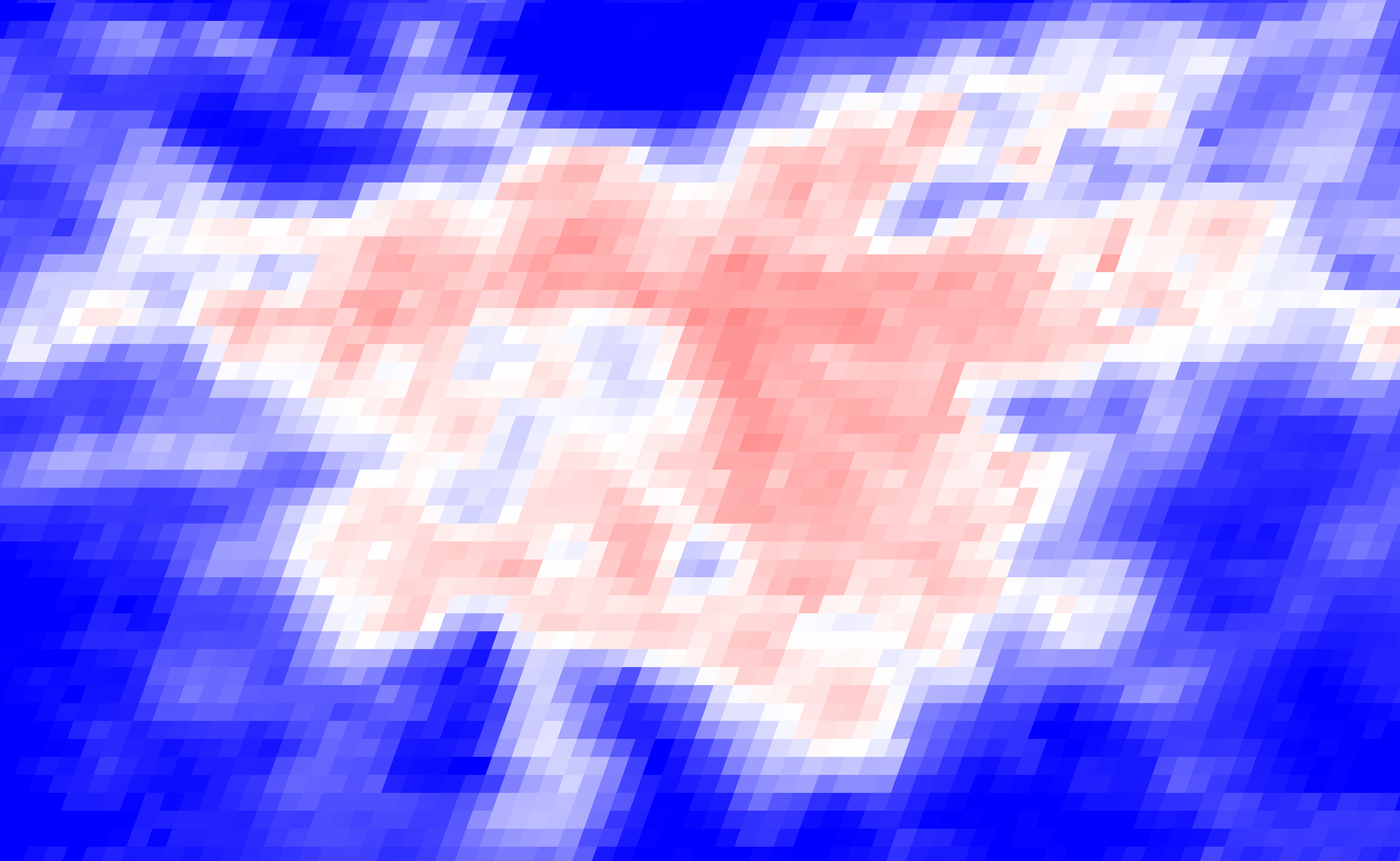}}
            \label{figureSaoPaulo2004}
        }%
\hspace{0.2 cm}
\subfigure[]{%
            
         \fbox{\includegraphics[height=2.2 cm]{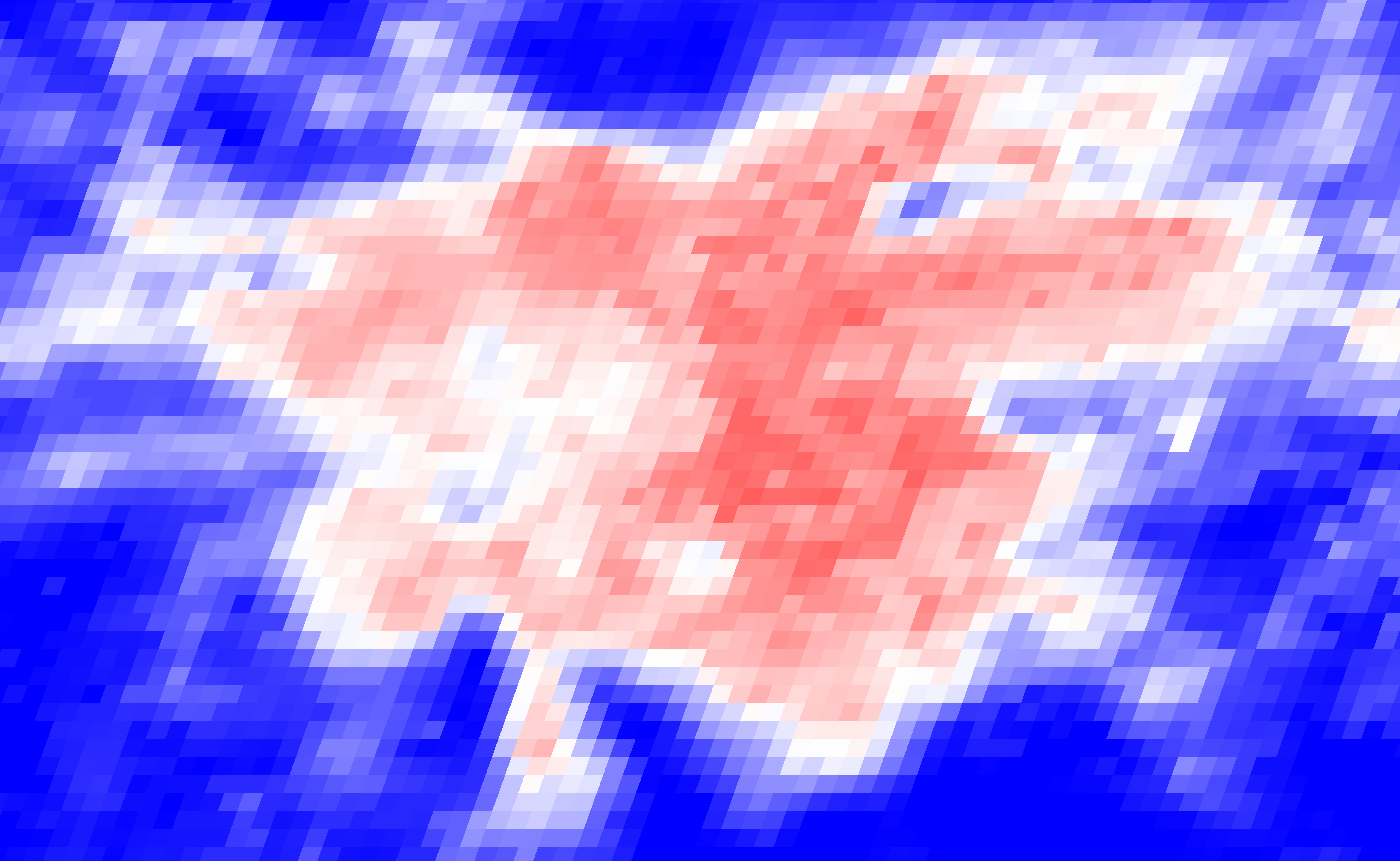}}
            \label{figureSaoPaulo2005}
        }%

\subfigure[]{%
            
         \fbox{\includegraphics[height=2.2 cm]{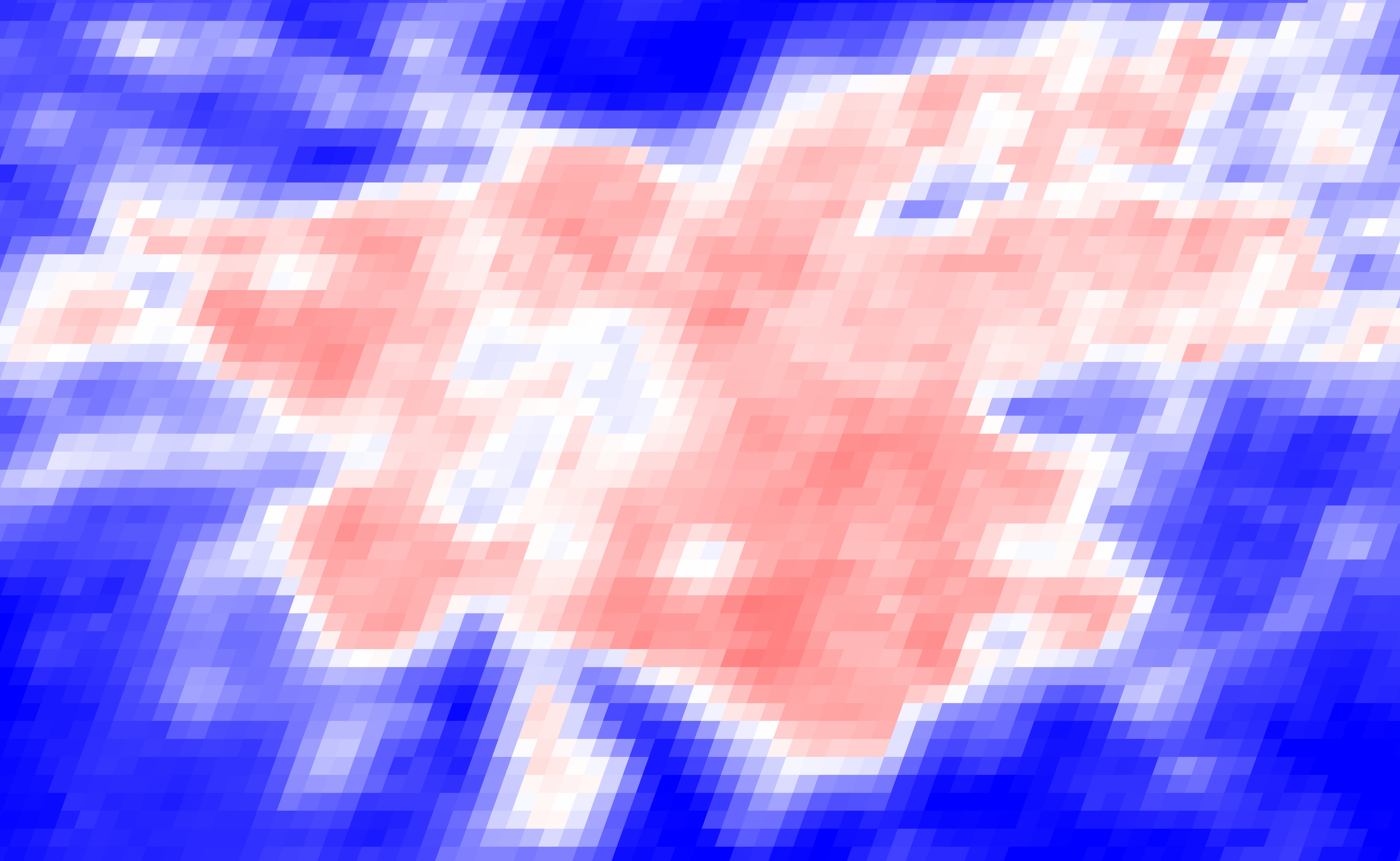}}
            \label{figureSaoPaulo2022}
        }%
\hspace{0.2 cm}
\subfigure[]{%
            
         \fbox{\includegraphics[height=2.2 cm]{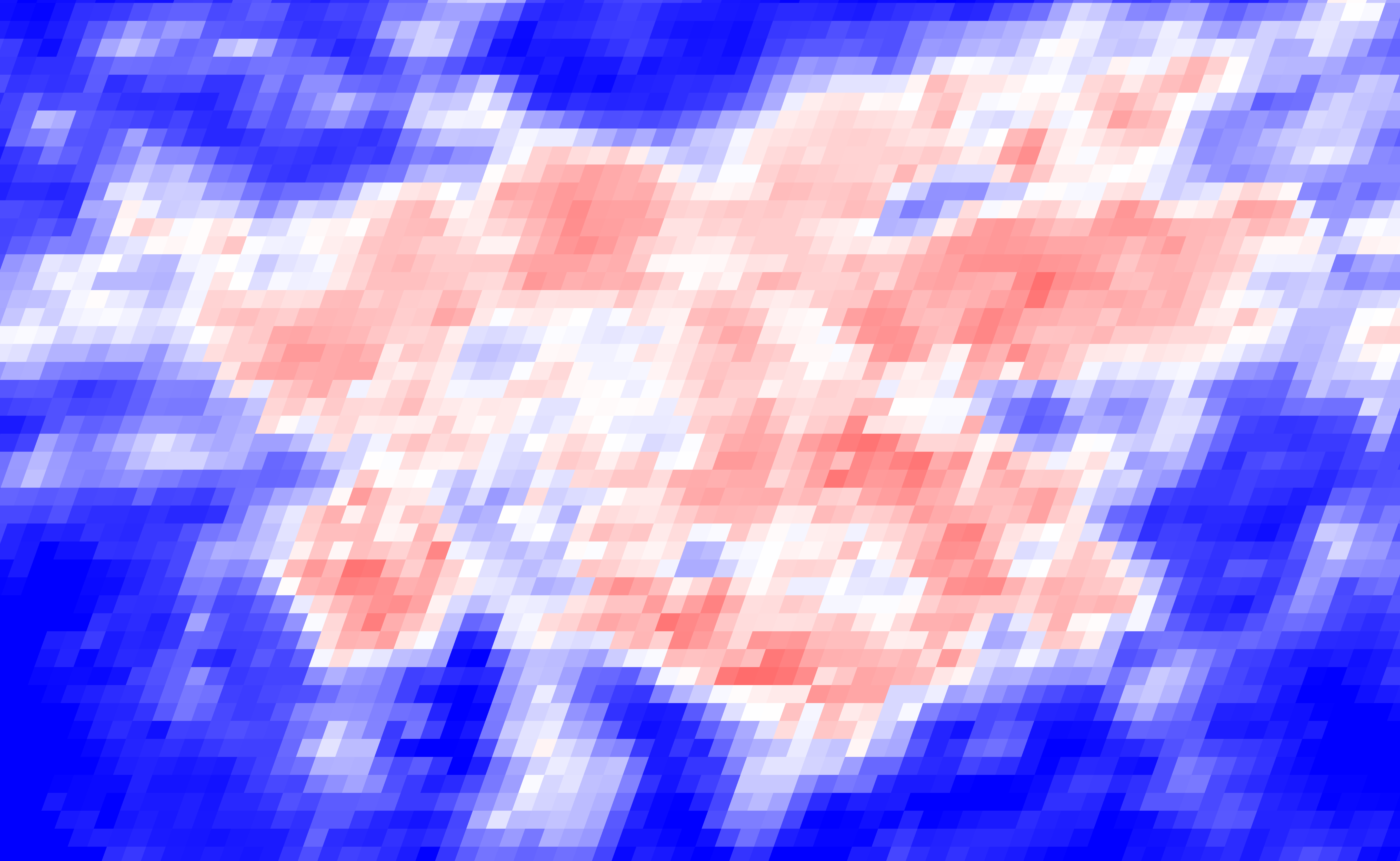}}
            \label{figureSaoPaulo2023}
        }%

\subfigure[]{%
            
         \fbox{\includegraphics[height=2.2 cm]{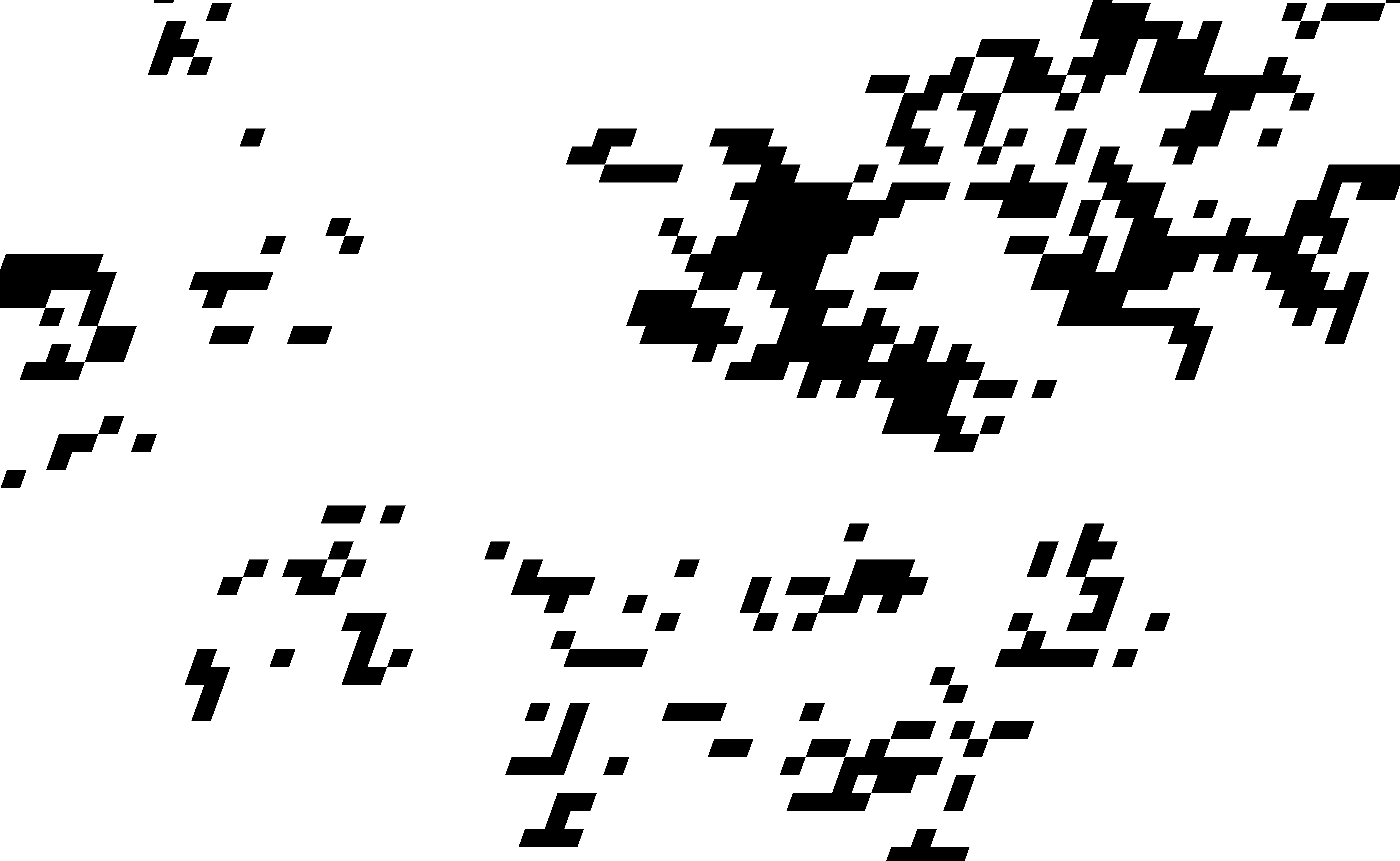}}
            \label{figureSaoPauloStd}
        }%
\hspace{0.2 cm}
\subfigure[]{%
            
         \fbox{\includegraphics[height=2.2 cm]{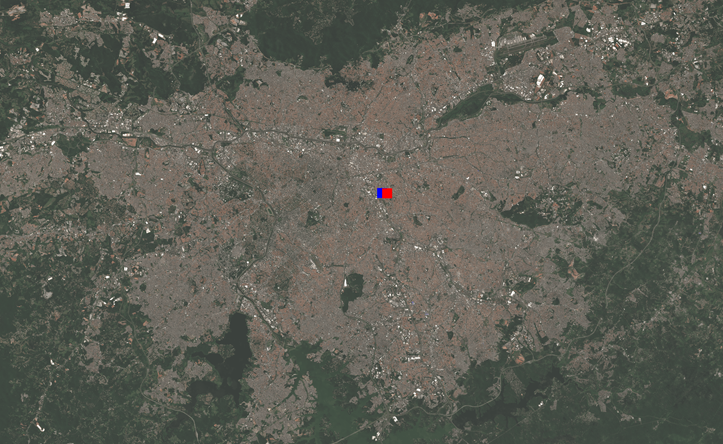}}
            \label{figureSaoPauloRGB}
        }%
\hspace{0.2 cm}
\subfigure[]{%
            
         \fbox{\includegraphics[height=2.2 cm]{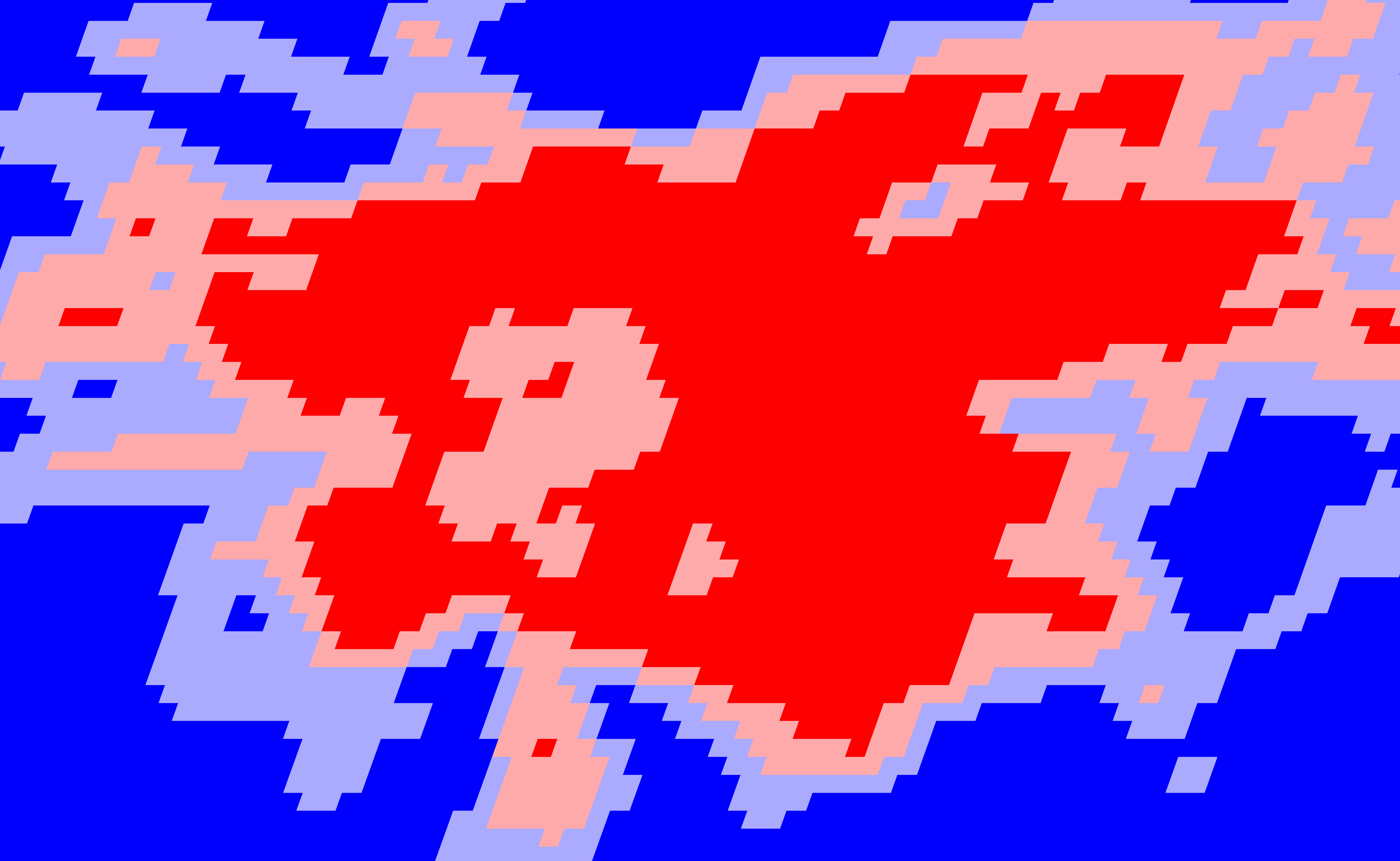}}
            \label{figureSaoPauloCluster}
        }%

\caption{Visualization of LST variation in a ROI around Sao Paulo, Brazil for year: (a) 2004, (b) 2005, (c)  2022, (d) 2023. (e) Areas with strong LST temporal variation shown in black. (f) RGB image of the analyzed scene with center of mass in 2004 and 2023 shown in blue and red, respectively. (g) clusters/groups obtained from LST time-series.}
\label{figureLSTSaoPaulo}
\end{figure}

\begin{figure}[!h]
\centering
\subfigure[]{%
            
         \fbox{\includegraphics[height=2.05 cm]{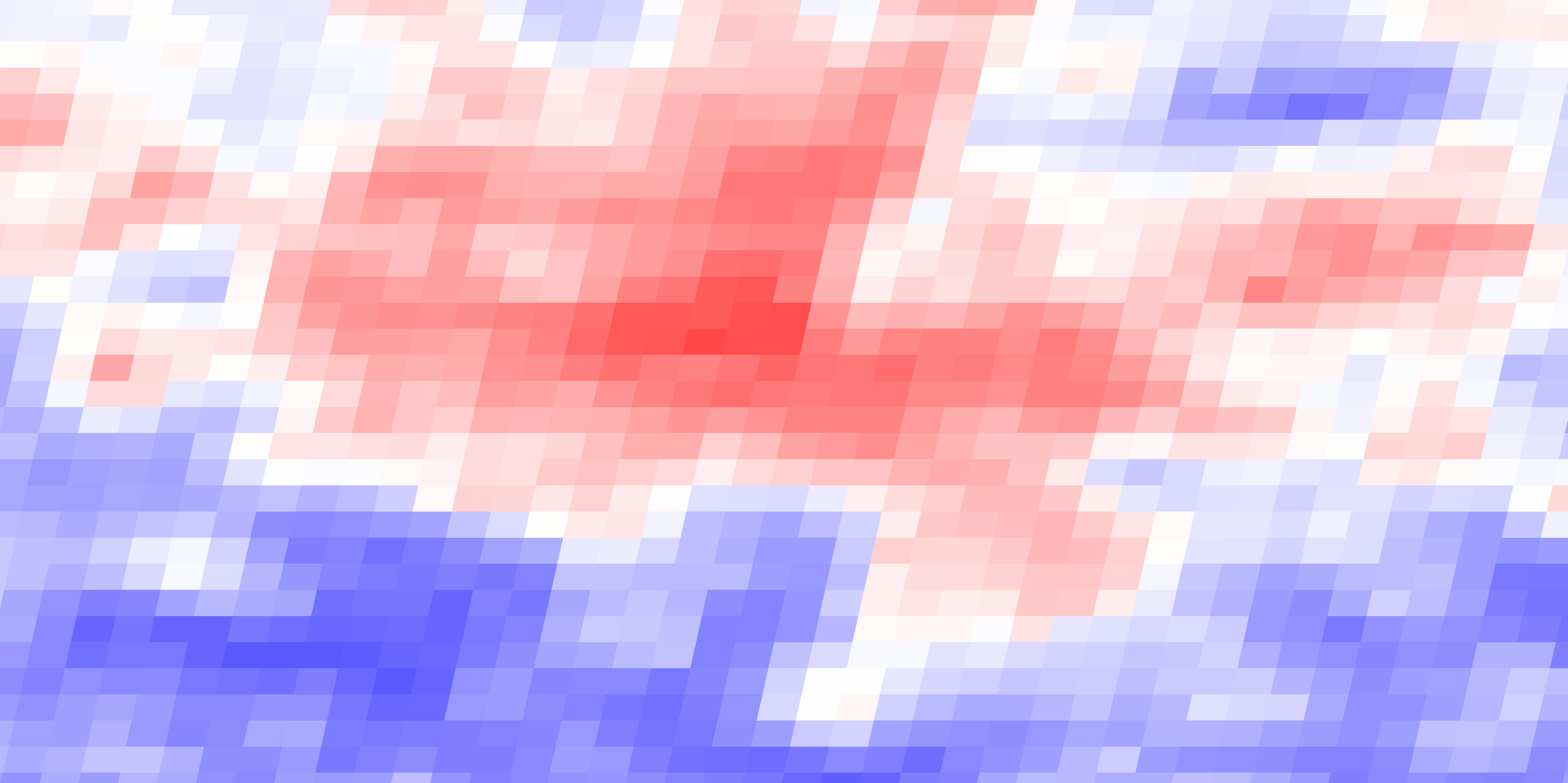}}
            \label{figureMunich2003}
        }%
\hspace{0.2 cm}
\subfigure[]{%
            
         \fbox{\includegraphics[height=2.05 cm]{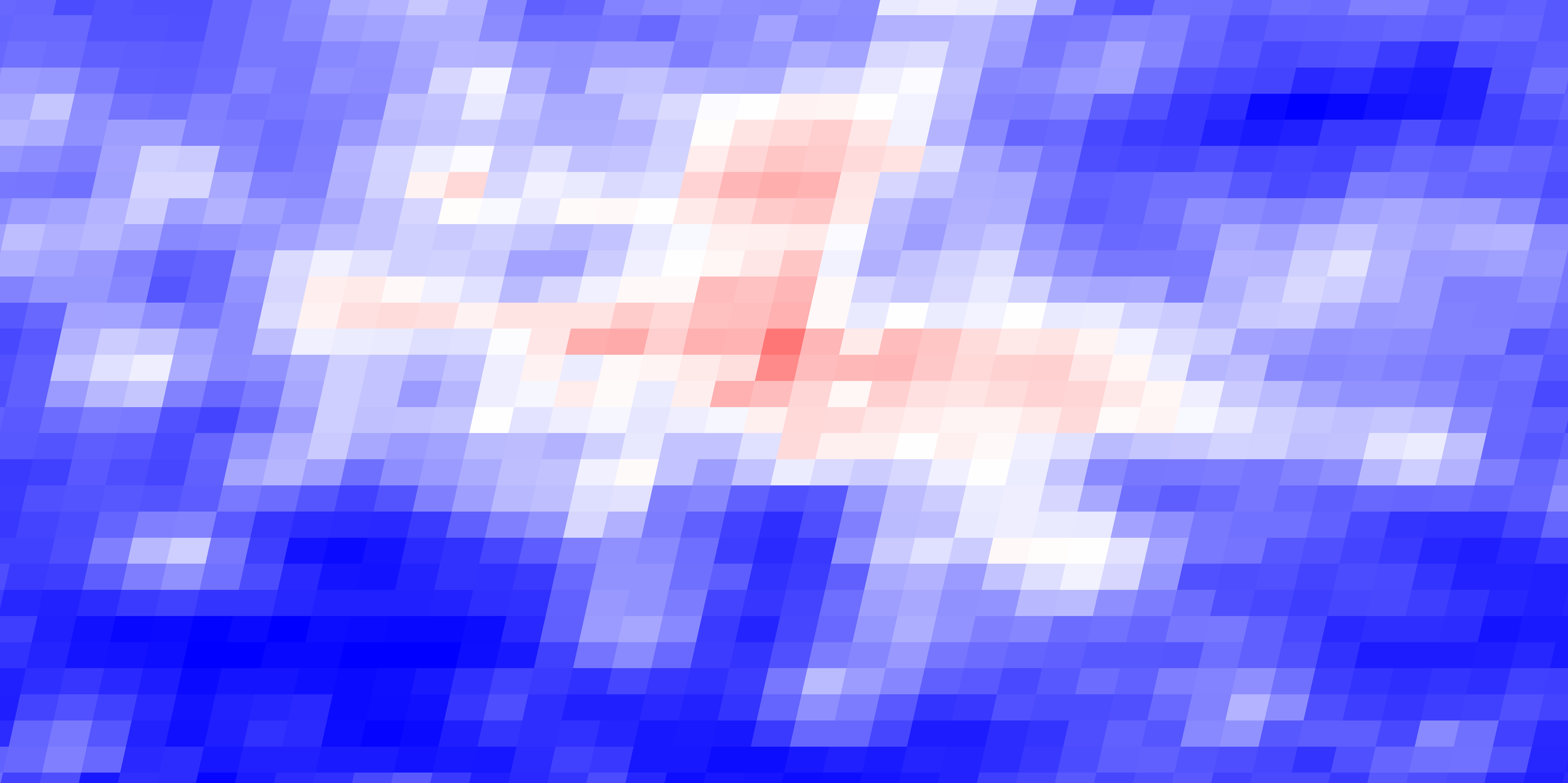}}
            \label{figureMunich2004}
        }%

\subfigure[]{%
            
         \fbox{\includegraphics[height=2.05 cm]{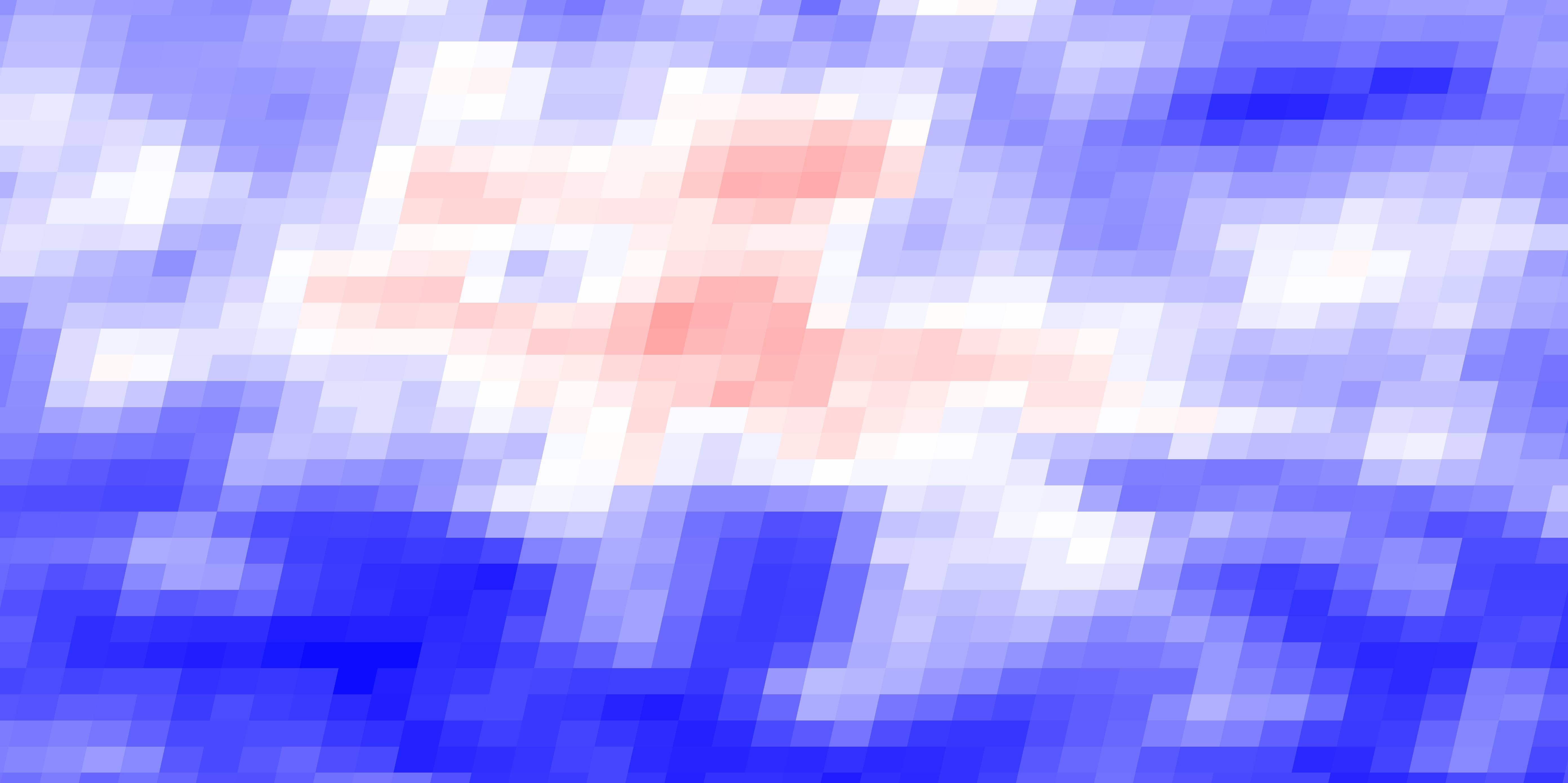}}
            \label{figureMunich2021}
        }%
\hspace{0.2 cm}
\subfigure[]{%
            
         \fbox{\includegraphics[height=2.05 cm]{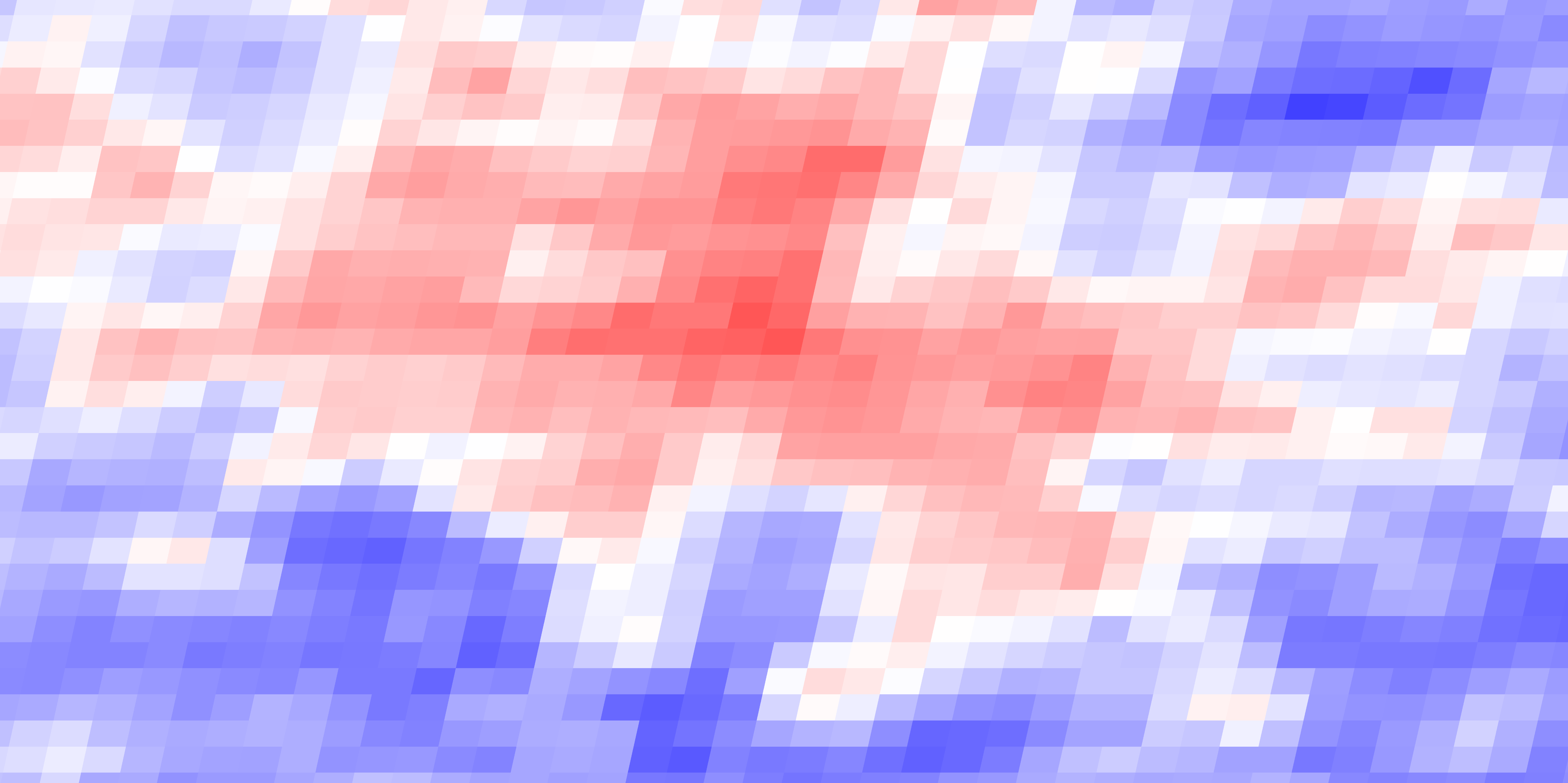}}
            \label{figureMunich2022}
        }%

\subfigure[]{%
            
         \fbox{\includegraphics[height=2.05 cm]{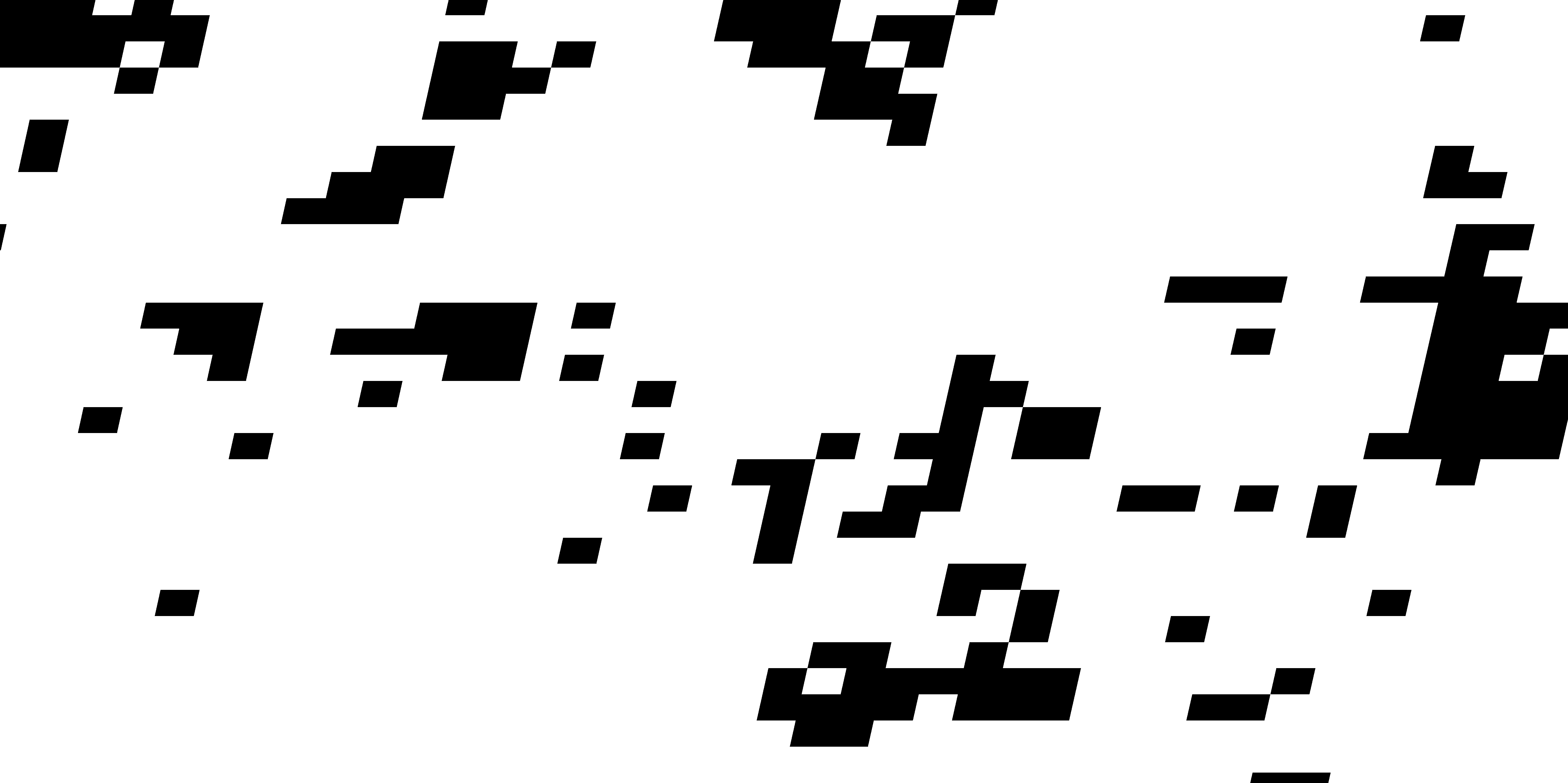}}
            \label{figureMunichStd}
        }%
\hspace{0.2 cm}
\subfigure[]{%
            
         \fbox{\includegraphics[height=2.05 cm]{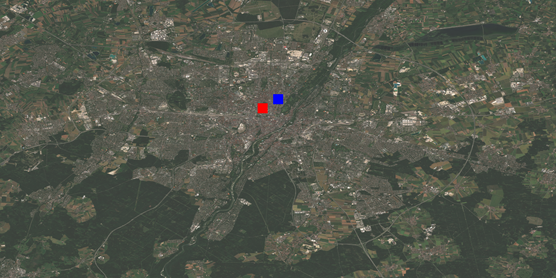}}
            \label{figureMunichRGB}
        }%

\subfigure[]{%
            
         \fbox{\includegraphics[height=2.05 cm]{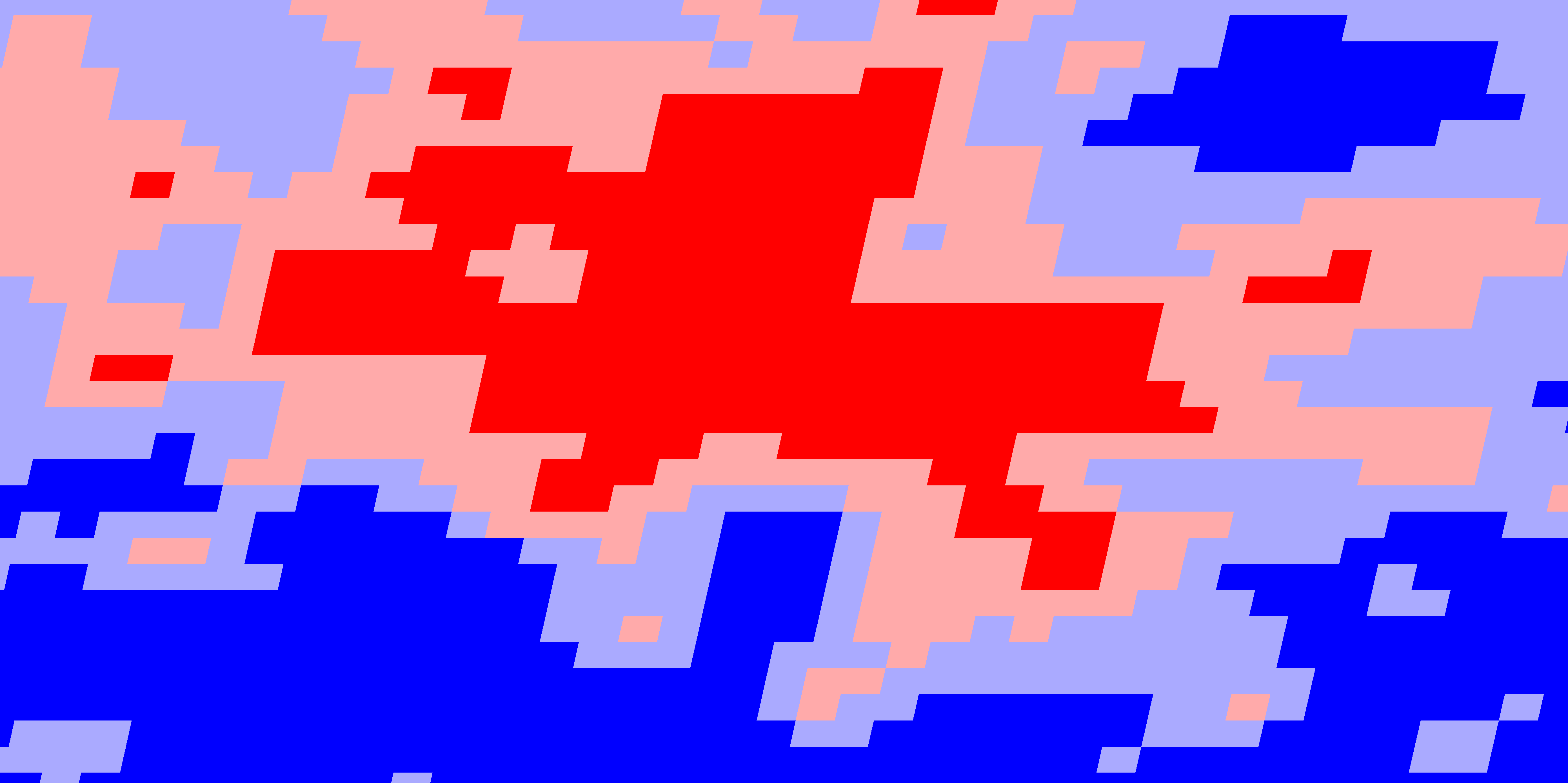}}
            \label{figureMunichCluster}
        }%

\caption{Visualization of LST variation in a ROI around Munich, Germany for year: (a) 2003, (b) 2004, (c)  2021, (d) 2022. e) Areas with strong LST temporal variation shown in black. (f) RGB image of the analyzed scene with center of mass in 2003 and 2022 shown in blue and red, respectively. (g) clusters/groups obtained from LST time-series.}
\label{figureLSTMunich}
\end{figure}

\section{Results}
\label{sectionExperimentalResult}

\subsection{Analyzed ROIs}
We conducted experiments on 20 years summer LST time-series on region of interests (ROIs) around Kolkata, Sao Paulo, and Munich:
\begin{itemize}
    \item The ROI around Kolkata is chosen between 88.202\textdegree \ and 88.502\textdegree  
 \ latitude and 22.464\textdegree \ and 22.764\textdegree \ longitude. We only consider data for the month of April and May, since these two months are typically hot without much rain. For each considered year, a median composite is formed. 20 years time-series is formed using data from 2004 to 2023.
 \item The ROI around Sao Paulo is chosen between -23.39\textdegree \ and -23.79\textdegree \ latitude and -46.94\textdegree \ and -46.29\textdegree \ longitude. We only consider data for the month of January and February, two of the warmest months in the region.  20 years time-series is formed using data from 2004 to 2023.
    \item The ROI around Munich is chosen between 11.332\textdegree \ and 11.832\textdegree \ latitude and 47.996\textdegree \ and 48.246\textdegree \ longitude. We only consider data for the month of June and July, two of the warmest months in Central Europe.  20 years time-series is formed using data from 2003 to 2022.
\end{itemize}
Two of the three analyzed cities are close to tropics and thus more susceptible to climate change \cite{bastin2019understanding}. Our ROIs are not the exact city boundaries, rather a region including the built-up region around the city.

\subsection{Result analysis}
\textbf{Kolkata} shows an increasing trend of LST over over the years. Figure \ref{figureLSTKolkata}
(a)-(c) show the LST for the years 2004, 2005, and 2006, respectively. Compared to them, we see observable difference for the years 2021, 2022, 2023 shown in Figures \ref{figureLSTKolkata}
(d)-(f). The LST has increased in the peripheral areas, more consistently in the center-east of the analyzed scene, which approximately corresponds to New Town and beyond, an area in greater Kolkata that has gone through significant infrastructure development in the last two decades.  Remarkably the increase in LST in this specific region is visible even for 2022, which otherwise shows less LST over the analyzed scene compared to its preceding and succeeding year. Furthermore, increment in LST is also observed for center-west and north-west. This consistent increase in LST is aligned with the fact that greater Kolkata is seeing large infrastructure development and conversion of more and more lands to built-up area. Figure \ref{figureKolkataStd} shows the areas with high LST variance over years. As expected, the suburban areas with more recent developments have LST variance. Figure \ref{figureKolkataRGB} shows  recent RGB image of the analyzed scene with the center of mass of LST in 2004 shown with a blue mark and the corresponding point for 2023 shown with a red mark. We observe upward and slightly eastward shift of the center of mass.
\par
The clustering obtained from LST time-series with number of clusters set as 4 is shown in Figure \ref{figureKolkataCluster}. We observe that one of the clusters closely resembles the main functional area of Kolkata and Howrah, in the two sides of river Ganga. Due to the presence of the river, this cluster (corresponding to highest LST) is spatially bifurcated. Another cluster shows the adjacent suburban areas, including those areas that are being developed more recently. The other two clusters represent other areas representing mostly non-built areas. Thus, we find that clusters obtained from the LST time-series automatically corresponds to the spatio-temporal pattern of the built areas. The IoU score computed between  two top LST classes and the building density map is 0.58.
\par
\textbf{Sao Paulo} also shows increasing trend of LST over the peripheral years, however less prominently than Kolkata.  Figure \ref{figureLSTSaoPaulo}
(a)-(d) show the LST for the years 2003, 2004, 2022, and 2023, respectively.  By plotting the standard deviation of LST time-series, this difference in the peripheral areas is prominent in Figure \ref{figureSaoPauloStd}. However,  overall the center of mass of the LST did not shift much, as shown in Figure
\ref{figureSaoPauloRGB}. This is also an indication that Sao Paulo might have had urban growth, however uniformly around the city. 
\par
Figure \ref{figureSaoPauloCluster} shows the clustering obtained from LST time-series. Remarkably, a
 part of the cluster corresponding to the second warmest zone lies inside the cluster corresponding to the warmest zone, thus showing a cooling within the city due to less prevalence of built structure. The IoU score computed between  two top LST classes and the building density map is 0.76, showing a high correlation between them.
\par
\textbf{Munich} shows more variation between successive years, however shows a relatively stable trend over longer period. Figure \ref{figureLSTMunich}
(a)-(b) show the LST for the years 2003 and 2004, respectively. We see a sharp difference between these two years. Similarly, such difference is observed between the years 2021 (Figure \ref{figureMunich2021}) and 2022 (Figure \ref{figureMunich2022}). However, when the year 2003 and 2022 are compared, there are negligible differences except a small region in south-east of the city. This is in consistent with the fact that the rate of new constructions in such a central European city is much less when compared to cities from developing countries such as Kolkata. Similar to the other two cities, we observe some peripheral areas showing more temporal variance of LST (Figure \ref{figureMunichStd}). The shift of center of mass of LST is negligible, as shown in Figure \ref{figureMunichRGB}.
\par
The clustering obtained from LST time-series is shown in Figure \ref{figureMunichCluster}. In this case we observe that the clusters almost form ring around each other, i.e., the warmest cluster lying in the center and less warmer clusters lying farther away from the center. The IoU score computed between  two top LST classes and the building density map is 0.59.

\section{Conclusions}
\label{sectionConclusion}
Along with rapid urbanization, our cities are getting impacted by increased warming in the summer. Specially the peripheral areas are getting more warmer than before. This paper showed this phenomenon w.r.t three cities from three continents. Furthermore, we made some interesting observations. In a city like Kolkata, where urban growth has been slightly tilted towards certain directions, the center of mass of LST shifted over time. The same does not hold for other two analyzed cities. Moreover a correlation between clusters obtained from the LST time-series and the building density is also found. In future work, we would like to consider other parameters, e.g., rainfall and air pollution indicators, in addition to LST while clustering. Furthermore, we would like to use deep learning based clustering, e.g., following \cite{saha2022unsupervised}. We would also like to formulate a generative approach, similar to \cite{diaconu2022understanding}, to investigate whether current RGB images of a ROI can be generated based on LST data and past RGB images. Furthermore, we would like to extend our study over more cities.

\bibliographystyle{ieeetr}
\bibliography{sigproc}

\end{document}